\documentclass[12pt]{article}
\usepackage{amsxtra,amssymb,amsthm,amsmath,latexsym}

\theoremstyle{plain}
\newtheorem{theorem}{Theorem}
\newtheorem*{theoremRamm}{Theorem (Ramm)}
\newtheorem*{remark}{Remark}

\newcommand{\refT}[1]{Theorem~\ref{T:#1}}
\newcommand{\refS}[1]{Section~\ref{S:#1}}

\def\barh{{\overline{h}}}
\def\nd{\noindent}

\def\bysame{\rule{.5in}{.005in},\ }

\def\ve{\varepsilon}

\def\lra{\longrightarrow}
\def\R{{\mathbb R}}

\def\C{{\mathbb C}}

\def\calF{{\mathcal F}}

\def\calM{{\mathcal M}}

\def\calQ{{\mathcal Q}}

\def\hatq{{\widehat{q}}}
\def\hatA{{\widehat{A}}}
\def\tildeq{{\widetilde q}}

\def\tildeD{{\widetilde D}}
\def\oH{{\overset{\circ}{H}}}
\def\oH1{{\overset{\circ}{H}\kern-.02in{}^1}}
\def\l{\ell}

\def\bee{\begin{equation*}}
\def\eee{\end{equation*}}
\def\be{\begin{equation}}
\def\ee{\end{equation}}

\begin{document}
\title{Distribution of particles which produces a "smart" material}

\author{A.G. Ramm\\
 Mathematics Department, Kansas State University, \\
 Manhattan, KS 66506-2602, USA\\
ramm@math.ksu.edu }

\date{}
\maketitle\thispagestyle{empty}

\begin{abstract} 

If $A_q(\beta, \alpha, k)$ is the scattering amplitude, corresponding to
a potential $q\in L^2(D)$, where $D\subset\R^3$ is a bounded domain,
and $e^{ik\alpha \cdot x}$ is the incident plane wave, then
we call the radiation pattern the function $A(\beta):=A_q(\beta, \alpha, 
k)$, where the unit vector $\alpha$,
the incident direction, is fixed, and $k>0$, the wavenumber, is fixed. 
It is shown that any function $f(\beta)\in L^2(S^2)$,
where $S^2$ is the unit sphere in $\R^3$, can be 
approximated with any desired accuracy by a radiation pattern:
$||f(\beta)-A(\beta)||_{L^2(S^2)}<\epsilon$, where $\epsilon>0$
is an arbitrary small fixed number. The potential $q$, corresponding to
$A(\beta)$, depends on $f$ and $\epsilon$, and can be 
calculated analytically.
There is a one-to-one correspondence between the above potential and
the density of the number of small
acoustically soft particles $D_m\subset
D$, $1\leq m\leq M$, distributed in an a priori given bounded domain
$D\subset\R^3$.  The geometrical shape of a small
particle $D_m$ is arbitrary, the boundary $S_m$ of $D_m$ is Lipschitz
uniformly with respect to $m$. The wave number $k$ and the direction
$\alpha$ of the incident upon $D$ plane wave are fixed.

It is shown that a suitable distribution of the above particles in
$D$ can produce the scattering amplitude $A(\alpha',\alpha)$,
$\alpha',\alpha\in S^2$, at a fixed $k>0$, arbitrarily close in the
norm of $L^2(S^2\times S^2)$ to an arbitrary given scattering amplitude
$f(\alpha',\alpha)$, corresponding to a real-valued potential $q\in 
L^2(D)$, i.e., corresponding to an arbitrary refraction coefficient in
$D$.

\end{abstract}

\section{Introduction}\label{S:1}

Let $D\subset\R^3$ be a bounded connected domain with 
Lipschitz boundary $S$.

The scattering of an acoustic plane wave 
$u_0=u_0(x)=e^{ik\alpha\cdot x}$, incident upon $D$, 
is described by the problem:
\be\label{e1}[\nabla^2+k^2n_0(x)]u=0\hbox{\quad in\quad} \R^3, \ee
\be\label{e2} u=u_0(x)+v,\ee
\be\label{e3}
 v=A(\alpha',\alpha)\ \frac{e^{ikr}}{r}+ o\left(\frac{1}{r}\right),
 \qquad r:=|x|\to\infty, \quad \frac{x}{r}:=\alpha'. \ee
The coefficient $A(\alpha',\alpha)$ is called the scattering 
amplitude, $k>0$ is the wave number, which is assumed fixed throughout 
the paper, and the dependence of $A$ on $k$ is not shown by this 
reason, $\alpha\in S^2$ is the direction of the incident plane 
wave, $\alpha'$ is the direction of the scattered wave, $n_0(x)$ 
is the known refraction coefficient in $D$, $n_0(x)=1$ in 
$D':=\R^3\setminus D$, and $v$ is the scattered field.

Let $D_m$, $1\leq m\leq M$, be a small particle, i.e., 
\be\label{e4}
  k_0a\ll 1, \hbox{\ where\ } a=\frac{1}{2}
  \max_{1\leq m\leq M} \,diam\, D_m,
\quad  k_0=k \max_{x\in D} |n_0(x)|. \ee
The geometrical shape of $D_m$ is arbitrary. We assume that 
$D_m$ is a Lipschitz domain uniformly with respect to $m$. This 
is a technical assumption which can be relaxed. It allows one
to use the properties of the electrostatic potentials. Denote
\be\label{e5}  d:=\min_{m\not= j}dist(D_m,D_j). \ee
Assume that 
\be\label{e6} a\ll d. \ee
 We do not assume that  $d\gg \lambda_0$, that is, that the distance between
the particles is much larger than the wavelength. Under our assumptions, it is possible that there are 
many small particles on the distances of the order of the wavelength.

The particles are assumed soft, i.e.,
\be\label{e7} u|_{S_m}=0\qquad 1\leq m\leq M. \ee
As a result of the distribution of many small particles in $D$, 
one obtains a new material, which we want to be a "smart" material, 
that is, a material which has some desired properties.
Specifically, we want this material to scatter the incident plane wave
according to an a priori given desired radiation pattern.
 Is this possible? If yes, how does one distribute the small particles
in order to create such a material?

We 
study this problem and solve the 
following two problems, which can be considered as problems of 
nanotechnology.

The first problem is:

\textit{
Given an arbitrary function $f(\beta)\in L^2(S^2)$, can one
distribute small particles in $D$ so that the resulting medium
generates the radiation pattern $A(\beta):= A(\beta,\alpha)$, at
a fixed $k>0$ and a fixed $\alpha\in S^2$, such that
\be\label{e8} \|f(\beta)-A(\beta)\|_{L^2(S^2)}\leq\ve, \ee 
where $\ve>0$ is an arbitrary small fixed number?
}

The answer is yes, and we give an algorithm for calculating 
such a distribution. This distribution is not uniquely defined 
by the function $f(\beta)$ and the number $\ve>0$.

The second problem is:

\textit{
Given a scattering amplitude $f(\alpha',\alpha)$, corresponding 
to some refraction coefficient $n(x)$ in a bounded domain $D$, 
can one distribute small particles in $D$ so that the resulting 
medium generates the scattering amplitude $A(\alpha',\alpha)$ 
such that
\be\label{e9}
  \|f(\alpha',\alpha)-A(\alpha',\alpha)\|_{L^2(S^2\times S^2)}
  \leq \ve, \ee
where $\ve>0$ is an arbitrary small fixed number?
}

The answer is yes, and we give an algorithm for calculating the 
density of the desired distribution of small particles given 
$f(\alpha',\alpha)$, $\forall\alpha',\alpha\in S^2$, $k>0$ being 
fixed.

To our knowledge the above two problems have not been studied in 
the literature. Our solution to these problems is based on some 
new results concerning the properties of the scattering 
amplitudes, on our earlier results on wave scattering 
by small bodies of arbitrary shapes (see \cite{R476}), and on our 
solution of the 3D inverse Schr\"odinger scattering problem 
with fixed-energy data \cite{R470},  \cite{R285}, 
\cite{R425}.

In \refS{2} we derive some new approximation properties of the 
scattering amplitudes. Essentially, we prove the existence of a 
potential $q\in L^2(D)$ such that the corresponding to this $q$ 
scattering amplitude $A(\beta)$, $\beta=\alpha'$, at an 
arbitrary fixed $\alpha\in S^2$ and an arbitrary fixed $k>0$,
approximates with any desired accuracy any given function 
$f(\beta)\in L^2(S^2)$ with a small norm. Moreover, we give formulas for 
calculating this $q$, and these formulas often work numerically for
$f$ which are not small.
The potential $q$ is related explicitly to a 
certain distribution of small particles in $D$. Consequently, we 
give formulas for calculating this distribution.

In \refS{3} we drive an equation describing the self-consistent 
field in the medium consisting of the small particles 
distributed in $D$. This equation is equivalent to a 
Schr\"odinger equation with a potential $q(x)$ supported in the 
bounded domain $D$ and related in a simple way to the density of 
the distribution of the small particles. 

The author has solved the 3D inverse scattering problem of 
finding a compactly supported potential $q$ from the knowledge 
of noisy fixed-energy scattering amplitude \cite{R425}. This 
algorithm allows one to calculate $q_\delta(x)$ from the knowledge of 
noisy data $f_\delta(\alpha',\alpha)$,  $\sup_{\alpha',\alpha\in S^2} 
|f_\delta-f|\leq \delta$, such that
\be\label{e10}
  \sup_{x\in D} |q_\delta(x)-q(x)|
  \leq\eta(\delta)\underset{\delta\to 0}{\lra} 0, \ee
where $q(x)$ is the  exact potential, generating the exact scattering 
amplitude $f(\alpha',\alpha)$ at a fixed $k>0$.
  
Applying this algorithm to the exact data $f(\alpha',\alpha)$ or to the
noisy data $f_\delta(\alpha',\alpha)$, one obtains a stable (in 
the sense \eqref{e10}) approximation of $q$, and, consequently, 
of the density of the distribution of small particles, which 
generates the scattering amplitude arbitrarily close to the a 
priori given scattering amplitude.

The author's solution of the 3D inverse scattering problem with 
the error estimates is presented in \refS{4}.

\section{Approximation properties of the scattering amplitudes}\label{S:2}

If $k>0$ is fixed, then the scattering problem 
\eqref{e1}--\eqref{e3} is equivalent to the Schro\"dinger 
scattering problem on the potential $q_0(x)$:
\be\label{e11}
  [\nabla^2+k^2-q_0(x)]u=0 \hbox{\quad in\ }\R^3, \ee
\be\label{e12}
  q_0(x)=\begin{cases}0\hbox{\ in\ }D',  \quad  D':=\R^3\setminus D&\\
  k^2[n_0(x)-1] \hbox{\qquad in\ }D.   \end{cases}\ee
The scattering solution $u= u_{q_0}$ solves (uniquely) the equation
\be\label{e13}
u_{q_0}=u_0-\int_D g(x,y) q_0(y)u_{q_0}(y)dy, 
  \quad g(x,y):=\frac{e^{ik|x-y|}}{4\pi|x-y|}. \ee
The corresponding scattering amplitude is:
\be\label{e14}
  A_0(\alpha',\alpha)=-\frac{1}{4\pi}
  \int_D e^{-ik\alpha'\cdot x} q_0(x)u_{q_0}(x,\alpha)dx,\ee
where the dependence on $k$ is dropped since $k>0$ is fixed.

If $q_0$ is known, then $A_0:=A_{q_0}$ is known.
Let $q\in L^2(D)$ be a potential and $A_q(\alpha',\alpha)$ be 
the corresponding scattering amplitude. Fix $\alpha\in S^2$ and 
denote
\be\label{e15}
  A(\beta):=A_q(\alpha',\alpha), \qquad \alpha'=\beta. \ee
Then
\be\label{e16}
  A(\beta)=-\frac{1}{4\pi}\int_D e^{-ik\beta\cdot x}h(x)dx,
  \qquad h(x):=q(x)u_q(x,\alpha). \ee

\begin{theorem}\label{T:1}
Let $f(\beta)\in L^2(S^2)$ be arbitrary. Then
\be\label{e17}
  \inf_{h\in L^2(D)} \bigg\| f(\beta)-
  \left( -\frac{1}{4\pi} \int_D e^{-ik\beta\cdot x}h(x)dx\right)
  \bigg\|  =0.\ee
\end{theorem}

\begin{proof}[Proof of \refT{1}]
If \eqref{e17} fails, then there is a function $f(\beta)\in 
L^2(S^2)$, $f\not=0$, such that
\be\label{e18}
  \int_{S^2} d\beta f(\beta) \int_D e^{-ik\beta\cdot x} 
  h(x)dx=0 \qquad \forall h\in L^2(D).\ee
This implies
\be\label{e19}
  \varphi(x):=\int_{S^2} d\beta f(\beta)e^{-ik\beta\cdot x}=0 
  \qquad \forall x\in D.\ee
The function $\varphi(x)$ is an entire function of $x$. 
Therefore \eqref{e19} implies
\be\label{e20}
  \varphi(x)=0\qquad \forall x\in \R^3.\ee
This and the injectivity of the Fourier transform imply 
$f(\beta)=0$. Note that $\varphi(x)$ is the Fourier transform 
of the distribution $f(\beta)\delta(k-\lambda)\lambda^{-2}$,
where $\delta(k-\lambda)$ is the delta-function and 
$\lambda\beta$ is the Fourier transform variable.
The injectivity of the Fourier transform implies 
$f(\beta)\delta(k-\lambda)=0$, so $f(\beta)=0$.

Theorem 1 is proved.
\end{proof}

Let us give an algorithm for calculating $h(x)$ in \eqref{e17} 
such that the left-hand side of \eqref{e17} does not exceed $\ve$, where 
$\ve>0$ is an arbitrary small given number.

Let $\{Y_\l(\beta)\}_{\l=0}^{\infty}$, $Y_\l=Y_{\l,m}$, $-\l\leq 
m\leq\l$, be the orthonormal in $L^2(S^2)$ spherical harmonics,
\be\label{e21}
  Y_{\l,m}(-\beta)=(-1)^\l Y_{\l,m}(\beta), \quad 
\overline{Y_{\l,m}(\beta)}=(-1)^{\l+m}Y_{\l,m}(\beta), \ee
\be\label{e22}
  j_\l(r):=\left(\frac{\pi}{2r}\right)^{1/2}
  J_{\l+\frac{1}{2}}(r), \ee
where $J_\l$ are the Bessel functions. It is known that
\be\label{e23}
  e^{-ik\beta\cdot x}=\sum_{\l=0,-\l\leq m\leq\l}
  4\pi(-i)^\l j_\l(kr) \overline{Y_{\l,m}(x^0)}
  Y_{\l,m}(\beta), \quad x^0:=\frac{x}{|x|}. \ee
Let us expand $f$ into the Fourier series with respect to spherical harmonics:
\be\label{e24}
  f(\beta)=\sum_{\l=0,-\l\leq m\leq\l} f_{\l,m} 
Y_{\l,m}(\beta).\ee
Choose $L$ such that
\be\label{e25}
  \sum_{\l>L}|f_{\l,m}|^2\leq \ve^2. \ee
With so fixed $L$, take $h_{\l,m}(r)$, $0\leq\l\leq L$, 
$-\l\leq m\leq\l$, such that
\be\label{e26}
  f_{\l,m}=-(-i)^\l \left(\frac{\pi}{2k}\right)^{1/2}
  \int^b_0 r^{3/2} J_{\l+\frac12}(kr) h_{\l,m}(r)dr,\ee
where $b>0$, the origin $O$ is inside $D$, the ball centered at the 
origin 
and of radius $b$ belongs to $D$, and $h_{\l,m}(r)=0$ for $r>b$.
There are many choices of $h_{\l,m}(r)$ which satisfy 
\eqref{e26}. If \eqref{e25} and \eqref{e26} hold, then the norm 
on the left-hand side of \eqref{e17} is $\leq\ve$.

A possible explicit choice of $h_{\l,m}(r)$ is
\be\label{e27}
  h_{\l,m}=
  \begin{cases}
  \frac{f_{\l,m}}{-(-i)^\l \sqrt{\frac{\pi}{2k}} 
     g_{1,\l+\frac12}(k)},& \l\leq L,\\
  0, & \l>L, \end{cases} \ee
where we have assumed that $b=1$ in \eqref{e26}, and used the 
following formula (see \cite[formula 8.5.8]{B}):
\be\label{e28}
  \begin{aligned}
  \int^1_0 x^{\mu+\frac12} J_\nu(kx)dx=k^{-\mu-\frac32}
  \left[\left(\nu +\mu-\frac12\right)
     kJ_\nu(r)S_{\mu-\frac12,\nu-1}(k) \right.\\
  \left.-k J_{\nu-1}(k)S_{\mu+\frac12,\nu}(k)+2^{\mu+\frac12}
  \frac{\Gamma\left(\frac{\mu+\nu}{2}+\frac34\right)}
       {\Gamma\left(\frac{\nu-\mu}{2}+\frac14\right)}\right]
  :=g_{\mu,\nu}(k),\end{aligned} \ee where $S_{\mu,\nu}(k)$ are Lommel's
functions, $\Gamma (x)$ is the Gamma-function, $h_{\l,m}(r)$
in \eqref{e27} do not depend on $r$, and we assume that $h(x)=0$ for 
$r:=|x|>1$. 

Let us prove that for any $q\in L^2$ there exists a $q\in L^2(D)$ such that
$q(x)u_q$ approximates $h(x)$ in $L^2(D)\hbox{-norm}$ with arbitrary
accuracy.

\begin{theorem}\label{T:2}
Let $h\in L^2(D)$ be arbitrary. Then
\be\label{e29}
  \inf_{q\in L^2(D)} \|h-qu_q(x,\alpha)\|=0.\ee
Here $\alpha\in S^2$ and $k>0$ are arbitrary, fixed.
There exists a potential $q$ such that $h=qu$, if $||h||_{L^2(D)}$
is sufficiently small. 
\end{theorem}

\begin{proof}[Proof of \refT{2}]
In this proof we first assume that the norm of 
$f$ is small, and then we 
drop the "smallness" assumption. If the norm
of $f$ is sufficiently small, the the norm of 
$h$ is small, so that
the condition 
\be\label{e30}
\inf_{x\in D}|u_0(x)-\int_Dg(x,y)h(y)dy|>0
\ee  is satisfied. 
Here $g$ is 
defined in formula \eqref{e13}. If this condition is satisfied,
then the formula  
\be\label{e31}
q(x)=h(x)[u_0(x)-\int_Dg(x,y)h(y)dy]^{-1}
\ee
yields the desired potential $q$.
The function $h$ generates the function $u:=\frac h q,$ where $u$ is the 
scattering solution, corresponding 
to the potential $q$, constructed by formula  \eqref{e31}.
Therefore, the infimum in \eqref{e29} is attained if condition 
\eqref{e30} is satisfied by the given $h$.

If $f$ is arbitrary, not necessarily samll, then $h$
is not necessarily small. If, nevertheless, condition \eqref{e30}
holds for this $h$, then the potential $q$, given by 
formula \eqref{e31}, belongs to $L^2(D)$ and yields the  
scattering amplitude $A_q(\beta)$ which satisfies \eqref{e8}.

On the other hand, if condition \eqref{e30} does not hold, then 
formula \eqref{e31} may yield a potential which is not locally 
integrable. In this case, as was proved in \cite{R517}, one can
perturb $h$ slightly, so that the perturbed $h$, denoted by $h_\ve$,
$||h-h_\ve||_{L^2(D)}<\ve,$  would yield by formula \eqref{e31}, with 
$h_{\ve}$ in place of $h$,
a potential $q_{\ve}\in L^2(D)$. This potential generates the scattering 
amplitude $A_{q_{\ve}}(\beta)$ which satisfies estimate \eqref{e8},
possibly with $c\ve$ in place of $\ve$, where the positive constant $c$ 
does not depend on $\ve$.
Theorem 2 is 
proved.
\end{proof}

Let us give a different point of view on the role of the "smallness" 
assumption. If $||q||_{L^2(D)}\to 0$, then the set of functions
$qu$ becomes a linear set. Thus, 
if \eqref{e29} fails, then there exists an $h\not=0$, 
$h\in L^2(D)$ such that
\be\label{e32}
  \int_D h(x)q(x)u_q(x,\alpha)dx=0 
  \qquad \forall q\in L^2(D) \quad ||q||<<1. \ee
Condition \eqref{e32} holds in the limit $||q||\to 0$ because
in this limit the set of functions $qu$ becomes linear.

Let $c=const>0$ be small. We will take $c\to 0$ 
eventually. Choose
$$  q=c\barh e^{-ik\alpha\cdot x}. $$
For sufficiently small $c>0$ the equation 
$$
u_q=e^{ik\alpha\cdot x}-\int_D g(x,y)c\barh 
e^{-ik\alpha\cdot y}u_qdy:=e^{ik\alpha\cdot x}-Tu_q $$
is uniquely solvable for $u_q$ in $C(D)$ because $\|T\|<1$ if 
$c>0$ is sufficiently small. We have
\be\label{e33}
  qu_q=q e^{ik\alpha\cdot x}-q Tu_q=c\barh+O(c^2),
  \quad c\to0.\ee
Substitute \eqref{e33} into \eqref{e32}, divide by $c$, and 
take $c\to0$. The result is:
\be\label{e34}  \int_D|h|^2dx=0. \ee
This implies $h=0$.

We describe the relation between $q(x)$ and the density 
distribution of small particles in \refS{3}. This relation 
makes it clear that a suitable distribution of small 
particles will produce any desirable potential $q\in 
L^2(D)$, and, consequently, any desirable scattering 
amplitude (radiation pattern) at an arbitrary fixed 
$\alpha\in S^2$ and $k>0$.

We describe the algorithm for calculating the above 
distribution of small particles, given $f(\beta)\in 
L^2(S^2)$, in \refS{3}.

\section{Scattering by many small particles.}\label{S:3}

The exact statement of the problem is:
\be\label{e35}
  [\nabla^2+k^2-q_0(x)]u=0 \quad \hbox{\ in\ }\R^3\setminus 
  \bigcup^M_{m-1} D_m,\ee
\be\label{e36}
  u=0\hbox{\ on\ }\bigcup^M_{m=1}S_m,\quad S_m:=\partial 
  D_m.\ee
\be\label{e37}
  u=e^{ik\alpha\cdot x} + v:=u_0+v, \ee
\be\label{e38}
  v=A(\alpha',\alpha)\frac{e^{ikr}}{r}+o\left(\frac1r\right),
  \quad r:=|\alpha|\to\infty, \quad \alpha'=\frac{x}{r}.\ee
We look for the solution of the form
\be\label{e39}
  u(x)=u_0(x)+\sum^M_{m=1}\int_{S_m}G(x,s)\sigma_m(s)ds,\ee
where $G(x,s)$ is the Green function which solves the 
scattering problem in the absence of small particles, i.e.:
\be\label{e40}
  [\nabla^2+k^2-q_0(x)]G=-\delta(x-y)\hbox{\ in\ }\R^3,\ee

\be\label{e41}
  \lim_{|x|\to\infty} |x| 
  \left(\frac{\partial G}{\partial|x|}-ikG\right)=0,\ee
and $u_0$ is the corresponding scattering solution.
It was proved in \cite{R190}, \cite{R470},  that
\be\label{e42}
  G(x,y)=\frac{e^{ik|x|}}{4\pi|x|} u_0(y,\alpha)
  +o\left(\frac{1}{|x|}\right), |x|\to\infty,
  \quad \alpha=-\frac{x}{|x|}, \ee
where $u_0$ is the scattering solution corresponding to 
$q_0$.

The function \eqref{e39} solves equation \eqref{e35}, 
satisfies the radiation condition \eqref{e38}, because
\be\label{e43}  u_0=u_0+v_0, \ee
where $v_0$ satisfies the radiation condition \eqref{e41}.
Therefore \eqref{e39} solves the problem 
\eqref{e35}--\eqref{e38} if $\sigma_m$ are such that the 
boundary condition \eqref{e36} is satisfied. All the above 
did not use the smallness of the particles.

Let us now use assumptions \eqref{e4} and \eqref{e6}.
Let $x_j\in D_j$ be an arbitrary point inside $D_j$.
Then
\be\label{e44}
  \sup_{S\in S_j} 
  |G(x,s)-G(x,x_j)\mid=O\left(ka+\frac{a}{d}\right),
  \quad |x-x_j|>d. \ee
This follows from the integral equation, relating $G$ and $g$:
\[ G(x,y)=g(x,y)-\int_{\R^3}g(x,z)q_0(z)G(z,y)dz,\]
 and from
the estimates:
\bee\begin{aligned} 
  \bigg| & \frac{e^{ik|x-s|}}{4\pi|x-s|}
  -\frac{e^{ik|x-x_j|}}{4\pi|x-x_j|}\bigg|
  = \frac{1}{4\pi|x-x_j|}
  \bigg|\frac{e^{ik(|x-s|-|x-x_j|)}|x-x_j|}{|x-s|}-1\bigg|, \\
  & k\bigg||x-s|-|x-x_j|\bigg|
  =k|x-x_j|\left(1+O\left(\frac{a}{d}\right)+O(ka)\right),\\
  &|x-s|=|x-x_j-(s-x_j)| 
  = |x-x_j|\left(1+O\left(\frac{a}{d}\right)\right).
  \end{aligned}\eee
From the integral equation for $G$ it follows that 
\[G(x,y)\sim g(x,y) \hbox{\quad as \quad} x\to y.\]
Therefore one may approximate \eqref{e39} as
\be\label{e45}
  u(x)=u_0(x)+\sum^M_{m=1} G(x,x_j) Q_m
  \left[1+O\left(ka+\frac{a}{d}\right)\right],\ee
where $|x-x_{j_m}|\geq d$ for all $m$, $1\leq m\leq M$, and
\be\label{e46}
  Q_m=\int_{S_m}\sigma_m(s)ds. \ee
Therefore, if one knows the numbers $Q_m$, $1\leq m\leq M$,
then one knows the scattering solution $u(x)$ at any point 
which is at a distance $\geq d$ from the nearest to $x$ 
small body.

Let us derive a linear algebraic system for calculating 
$Q_m$. To do this, let us use the boundary condition 
\eqref{e36}. We have:
\be\label{e47}
  \int_{S_m} G(s,t)\sigma_m dt
  =-\left[u_0(x_m)+\sum_{j\not=m}G(x_m,x_j)Q_j\right].\ee
Since $k|s-t|\leq2a\ll 1$, one has
\be\label{e48}
  G(s,t)\approx \frac{e^{ik|s-t|}}{4\pi|s-t|}
  =\frac{1}{4\pi|s-t|} \left(1+O(ka)\right). \ee
Consequently, equation \eqref{e47} can be written as
\be\label{e49}
  \int_{S_m} \frac{\sigma_m (t)}{4\pi|s-t|}dt 
   =-\left[u_0(x_m)+\sum_{j\not=m}G(x_m,x_j)Q_j\right].\ee
This is an equation for the electrostatic charge distribution $\sigma_m$ 
on the surface $S_m$
of the perfect conductor $D_m$,  charged to the potential 
which is 
given by  the right-hand side of \eqref{e49}.
The total charge on the surface of the conductor is given by the 
formula:
\be\label{e50}
  \int_{S_m} \sigma_m dt=Q_m=-C_m
  \left[u_0(x_m)+\sum_{j\not= m} G(x_m,x_j)Q_j\right],
  \quad 1\leq m\leq M, \ee
where $C_m$ is the electrical capacitance of the conductor 
$D_m$.
Equation \eqref{e50} is a linear algebraic system for the 
unknown $Q_j$.

Assume that the distribution of small bodies $D_m$ in $D$ 
is such that
\be\label{e51}
  \lim_{M\to\infty} \sum_{D_m\subset\tildeD} C_m
  =\int_\tildeD C(x)dx, \ee
where $\tildeD$ is an arbitrary subdomain of $D$. This means that 
$ C(x)$ is the limiting density of the capacitance per unit volume around 
an arbitrary point $x\in D$.
Then the relation \eqref{e45} in the limit \[M\to\infty, \quad ka\to 
0,\quad 
\frac{a}{d}\to 0,\] takes the form
\be\label{e52}
  u(x)=u_0(x)-\int_D G(x,y)C(y) u(y)dy,\ee
where $C(x)$ is defined in \eqref{e51}. 
An equation which 
is similar to \eqref{e52}, with $g(x_j,y)$ in place of 
$G(x_j,y)$, has been derived in \cite{MK} by a different 
argument.

Equation \eqref{e52} is equivalent to the Schr\"odinger 
equation
\be\label{e53}
  \left[\nabla^2+k^2-q_0(x)-C(x)\right]u=0,\ee
and $u(x)$ is the scattering solution corresponding to the 
potential 
\be\label{e54}  q(x)=q_0(x)+C(x).  \ee
If $q_0(x)$ is known (which we assume), then $q(x)$ and 
$C(x)$ are in one-to-one correspondence.

If the small particles $D_m$ are identical, and $C_0$ is the 
electrical capacitance of a single particle, then
\be\label{e55}  C(x)=N(x)C_0, \ee
where $N(x)$ is the density of the number of particles
in a neighborhood of the point $x$, that is, the number of particles per 
unit volume around point $x$.

Therefore, given $f(\beta)\in L^2(S)$, one finds $q(x)$, 
such that $\|A_q(\beta)-f(\beta)\|\leq\ve$, where $A_q(\beta)$ is
the scattering amplitude, corresponding to the potential $q$,
the energy $k^2>0$ and the incident direction $\alpha$ being fixed,
and $\beta=\alpha'$ is the direction of the scattered wave.

Let us describe the steps of our algorithm.\par

\smallskip
\nd\underbar{Step 1.} Given $f(\beta)$, find $h\in 
L^2(D)$.\par

\smallskip
\nd\underbar{Step 2.} Given $h\in L^2(D)$, find $q$ such 
that $\|h-q(x)u_q(x)\|_{L^2(D)}\leq\ve$.\par

\smallskip
Let us elaborate on Step 2. First, assume the
existence of a potential $q$, such that $h=qu$. Consider the equation 
\be\label{e56}
  u=u_0-\int g q u_q dy=u_0-\int_D gh dy. \ee
We have \[q u_q:=h.\] Thus,
\be\label{e57}
  A_q(\beta)=-\frac{1}{4\pi}\int_D e^{-ik\beta\cdot x} 
  h(x)dx. \ee
Multiply \eqref{e56} by $q$. Then
\bee  h=u_0q-q\int_D ghdy. \eee
Therefore, if 
\[\inf_{x\in D}|u_0(x)-\int_D g(x,s) h(y)dy|>0,\]
then the solution of the equation $qu_q=h$ is unique and is 
given by the formula:
\be\label{e58}
  q(x)=\frac{h(x)}{u_0(x)-\int_D ghdy}. \ee
Formula \eqref{e58} yields a potential for which 
$A_q(\beta)$ is given by formula \eqref{e57}, and the 
corresponding scattering solution is given by formula 
\eqref{e56}. All this is true provided, for example, that
\be\label{e59}
  \sup_{x\in D} \bigg|\int_D g(x,y)h(y)dy\bigg|<1.\ee
Inequality \eqref{e59} holds  if $h$ is fixed and $diam\,D$ is 
sufficiently small, because of the following estimate: 
\[\sup_x|\int_D g(x,y)h(y)dy|\leq (4\pi)^{-\frac 1 2} \|h\|_{L^2(D)} 
(diam\, D)^{\frac 1 2}.\]
Inequality \eqref{e59} also holds  if $||h||_{L^2(D)}$ is
sufficiently small and $D$ is fixed. The norm $||h||_{L^2(D)}$
is small if $||f||_{L^2(S^2)}$ is sufficiently small.
For the formula \eqref{e58} to yield the desired potential, the 
inequality \eqref{e59} is not necessary.
If one can  find a 
potential $q(x)$ from the given $h$ by formula \eqref{e58},
then
this $q$ generates the scattering solution by the formula
\be\label{e60}   u_q=u_0-\int_D ghdy \ee
and 
\be\label{e61}  h=q(x)u_q(x).  \ee
The potential $q$ can be found by formula \eqref{e58},
provided that $f(\beta)$ is sufficiently small, because then $h$
will be sufficiently small as follows, e.g., from \eqref{e27}.

If $q$ is found, then
\be\label{e62}   N(x)=\frac{q(x)-q_0(x)}{C_0}. \ee
Thus, the corresponding distribution density of small 
particles is given analytically.

Analytical formulas, which allow one to calculate $C_0$ with 
any desired accuracy, are derived in \cite{R476}, see also formula (91) 
below.

\begin{remark}
If $f(\beta)$ corresponds to a real-valued $q(x)$, then 
formula \eqref{e58} yields a real-valued potential.
In general, formula \eqref{e58} yields a complex-valued 
potential. To get a complex-valued potential by a formula, 
similar to \eqref{e55}, one has to replace the boundary 
condition \eqref{e36} by the condition
\be\label{e63}
  u_N=\zeta u\hbox{\quad on\quad} S_m, \ee
where $N$ is the exterior unit normal to the boundary $S$, 
and $\zeta$ is 
a complex constant, the impedance.
Then $C_0$ in \eqref{e55} should be replaced by the quantity:
\be\label{e64}
  C_\zeta =\frac{C_0}{1+\frac{C_0}{ \zeta S}}, \ee
(see \cite{R476}), and, therefore, formula \eqref{e55} 
yields a complex-valued $C_\zeta (x)$ if $\zeta$ is a complex number.
\end{remark}

Suppose that a given $h$ corresponds to a potential $q(x)\in L^2(D)$
in the sense that $h=q(x)u(x),$ where $u(x)$ is the scattering solution
corresponding to this $q(x)$ at the wavenumber $k>0$ and with the incident
direction $\alpha$. Then formula (58) defines $q(x)$, and the
corresponding
scattering solution is $u=\frac {h(x)}{q(x)}$.

If formula (58) does not produce a $p\in L^2(D)$, then one can replace
$h$ in (58) by an $h_\ve$, $||h-h_\ve||_{L^2(D)}<\ve$,
and get a square-integrable
potential $q_\ve$ by formula (58) with  $h$ replaced by $h_\ve$.
If $\ve$ is sufficiently small, this
potential $q_\ve$ generates the radiation pattern, which
differs by $O(\ve)$ from the desired $f$.

\section{Ramm's solution of the 3D inverse scattering problem with 
fixed-energy data}\label{S:4}

We follow \cite{R425}, \cite{R470}.
Consider first the inversion of the exact data 
$A_q(\alpha',\alpha)$. Let
\be\label{e65}
  A_q(\alpha',\alpha)=\sum^\infty_{\l,\l'=0}
  A_{\l,m,\l',m'} Y_{\l',m'}(\alpha')Y_{\l,m}(\alpha). \ee
It is proved in \cite{R470} that
\be\label{e66}
   \left|Y_\l (\theta)\right| \leq \frac{1}{\sqrt{4 \pi}} 
        \frac{e^{r |Im \theta|}}{|j_\l (r)|}, \quad
        \forall r > 0, \quad \theta \in \calM,
\ee
where
\bee
 \calM=\{z:z\in \C^3, z\cdot z=k^2\},
 \quad z\cdot\zeta:=\sum_{j=1}^3 z_j\zeta_j.\eee
Estimate (66) allows one to prove ( \cite{R425}) that
the series (65) converges absolutely for $\alpha'=\theta'\in \calM$,
so that the exact data $ A_q(\alpha',\alpha)$ allow one to calculate 
the values  $ A_q(\theta',\alpha)$, $\theta'\in \calM$. These values are
used below in the inversion formula (68).

One can prove \cite{R470}, that any $\xi\in\R^3$ can be 
written (nonuniquely) as
\be\label{e67}
  \xi=\theta'-\theta, \quad \theta',\theta\in\calM,
  \quad |\theta|\to\infty. \ee
In \cite{R470} explicit analytical formulas are given for 
$\theta'$ and $\theta$ satisfying \eqref{e67}.

The exact data $A(\alpha',\alpha)$ admit an analytic 
continuation from $S^2\times S^2$ onto $\calM\times S^2$.
Let 
\bee \tildeq(\xi):=\int_D q(x) e^{-i\xi\cdot x}dx. \eee
The inversion formula, proved in \cite{R470}, is
\be\label{e68}
\tildeq(\xi)=\lim_{\substack{|q|\to\infty\\ \theta'-\theta=\xi, 
\theta',\theta\in\calM}}
 \left[-4\pi \int_{S^2} 
A(\theta',\alpha)\nu(\alpha,\theta)d\alpha\right], \ee
where \eqref{e67} holds and $\nu (\alpha,\theta)$ is 
an arbitrary approximate solution to the problem
\be\label{e69}
  \calF(\nu):=\int_{a_1\leq|x|\leq b} 
  |\rho(x)|^2dx=inf:=d(\theta).\ee
Here
\be\label{e70}
  \rho(x):=e^{-i\theta \cdot x}
  \int_{S^2}u(x,\alpha)\nu(\alpha,\theta)d\alpha-1,\ee
$a_1>0$ is a radius of a ball which contains $D$ as a 
strictly inner-subdomain, and $b>a_1$ is an arbitrary 
fixed number. The approximate solution $\nu$ to 
\eqref{e69} is understood in the following sense:
\be\label{e71}
  \calF(\nu)\leq 2d(\theta). \ee
This means that it is not necessary to find a very accurate approximation 
of the infimum in problem (69). It is sufficient, for example, to find 
any function $\nu (\alpha, \theta)$ for which the functional (69)
takes the value not more than $2d(\theta)$, and the inversion formula
(68) holds with such  $\nu (\alpha, \theta)$. Also, formula (73) below,
with the error term, holds as well.

It is proved in \cite{R470} that
\be\label{e72}
  d(\theta)\leq\frac{c}{|\theta|},\qquad\theta\in\calM,\ee
where $c=c(||q||)>0$ is a constant depending on an 
$L^\infty(D)$ norm of $q$.
Therefore, given the exact data $A_q(\alpha',\alpha)$, one 
recovers the potential $q(x)$ by formula \eqref{e68}.

The error estimate of formula \eqref{e68} is given by the 
formula:
\be\label{e73}
  \tildeq(\xi)=-4\pi\int_{S^2}A(\theta',\alpha)\nu(\alpha,\theta)
  d\alpha + O\left(\frac{1}{|\theta|}\right),
  \quad |\theta|\to\infty, \ee
where \eqref{e67} holds. 

If $q(x)$ is found, then
\be\label{e74}
  N(x) C_0=q(x)-q_0(x),  \ee
so that the density of distributions of small particles is 
found explicitly.

Consider now inversion of noisy data $A_\delta(\alpha',\alpha)$,
\be\label{e75}
  \sup_{\alpha',\alpha\in S^2}
  |A_\delta(\alpha',\alpha)-A(\alpha',\alpha)|\leq\delta.\ee
Here $A(\alpha',\alpha)$ corresponds to an exact potential, 
and $A(\alpha',\alpha)$ is not known. Instead, its noisy 
measurements $A_\delta(\alpha',\alpha)$ are known.

Define
\be\label{e76}
  N(\delta)=\left[\frac{|\ln\delta|}{\ln|\ln\delta|}\right],\ee
where $[x]$ is the integer nearest to $x>0$,
\be\label{e77}
  \hatA_\delta(\theta',\alpha)
  =\sum^{N(\delta)}_{\l=0} A_{\delta\l}(\alpha)Y_\l(\theta'),
  \quad \sum_\l=\sum_\l \sum_{-\l\leq m\leq\l}, \ee
\be\label{e78}
  u_\delta(x,\alpha)=e^{ik\alpha\cdot x}
  +\sum^{N(\delta)}_{\l=0} A_{\delta\l}(\alpha)Y_\l(\alpha')
  h_\l(kr), \quad \alpha':=\frac{x}{r}, \quad r=|x|, \ee
\be\label{e79}
  \rho_\delta(x;\nu)=e^{-i\theta\cdot x}
  \int_{S^2}u_\delta (x,\alpha)\nu(\alpha)d\alpha-1,
  \quad \theta\in\calM,\ee
\be\label{e80}
  \mu(\delta)=e^{-\gamma N(\delta)}, 
  \qquad \gamma=\ln \frac{a_1}{b_0}>0, \ee
\be\label{e81}
  b_0:=\frac12 \,diam\, D, \qquad \kappa=|Im\theta|. \ee
Let
\be\label{e82}
  b_0<a_1<b , \ee
where $a_1$ and $b$ are arbitrary positive fixed numbers.
Consider the problem:
\be\label{e83}
  |\theta|=\sup:=\vartheta(\delta) \ee
under the constraints
\be\label{e84}
  |\theta|\left[\|\rho_\delta(\nu)\|_{L^2(\{x: a_1\leq|x|\leq b\})}
  + \|\nu\|_{L^2(S^2)}
  e^{\kappa b} \mu(\delta)\right] \leq c, \ee
\be\label{e85}
  \theta\in\calM, \quad\theta'-\theta=\xi, 
  \quad\theta',\theta\in\calM,\ee 
where $c>0$ is a sufficiently large constant, and $b_0<a_1<b$.

It is proved in \cite{R470} that
\be\label{e86}
  \vartheta(\delta)=O
  \left(\frac{|\ln\delta|}{\ln|\ln\delta|)^2}\right)
  \qquad \delta\to 0.\ee
Let $\theta(\delta)$ and $\nu_\delta(\alpha)$ be any 
approximate solution to \eqref{e83}--\eqref{e85} in the sense 
that 
\be\label{e87}
  |\theta(\delta)|\geq\frac12\vartheta(\delta). \ee
Define
\be\label{e88}
  \hatq_\delta:=-4\pi\int_{S^2}A_\delta(\theta',\alpha)
  \nu_\delta(\alpha)d\alpha.\ee
The following result is proved in \cite{R470}

\begin{theoremRamm}
One has
\be\label{e89}
  \sup_{\xi\in\R^3}|\hatq_\delta-\tildeq(\xi)|
  =O\left(\frac{(\ln|\ln\delta|)^2}{|\ln\delta|}\right),  
  \qquad \delta\to 0.\ee
\end{theoremRamm}

This result gives an inversion formula for finding the potential 
from noisy fixed-energy scattering data.

Thus, the algorithm for finding the density of the distribution 
of small particles from the fixed-energy scattering data 
$A(\alpha',\alpha)$ can be formulated as follows:\par

\smallskip
\nd\underbar{Step 1}. 
Given $A(\alpha',\alpha)$, find $q(x)$ using the inversion 
formulas \eqref{e68} in the case of the exact data or 
\eqref{e88} in the case of noisy data.\par

\smallskip
\nd\underbar{Step 2}. 
Find the density of the distribution of the small particles by 
formula \eqref{e62}, where formulas for $C_0$ are given in 
\cite{R470}.
\be\label{e90}
  |C_0-C^{(n)}|=O(\calQ^n), \qquad 0<\calQ <1,\ee
where $\calQ$ depends only on the geometry of the surface,
\be\label{e91}
  C^{(n)}=4\pi|S|^2
  \left\{ \frac{(-1)^n}{(2\pi)^n}
  \int_S\int_S \frac{dsdt}{r_{st}}
  \int_S\underset{n\,\, integrals }{\dots}\int_S
  \psi(t,t_1)\dots \psi(t_{n_1},t_n)dt_1\cdot 
dt_n\right\}\ee
\be\label{e92}
  \psi(t,s)=\frac{\partial}{\partial N_t}\,\frac{1}{r_{st}},
  \quad r_{st}=|s-t|,
  \quad |S|=\,meas\, S, \ee
$S$ is the surface of the conductor, $C_0$ is the 
electrical capacitance of this conductor, and $N_t$ is the exterior normal 
to 
$S$ at the point $t$.

In particular, for $n=0$ one gets
\be\label{e93}
  C^{(0)}=\frac{4\pi|S|^2}{J},
  \qquad J:=\int_S\int_S \frac{dsdt}{r_{st}}.\ee
It is proved in \cite{R476} that
\be\label{e94}
  C^{(0)}\leq C_0.\ee
Formula \eqref{e91} given an approximate value $C^{(n)}$ of 
the electrical capacitance of a perfect conductor placed in 
the space with dielectric permittivity $\ve_0=1$. If  $\ve_0\neq 1$,
then  one 
has to 
multiply the right-hand side of \eqref{e91} by $\ve_0$.

\end{document}